\documentclass[lettersize,journal]{IEEEtran}
\usepackage{amsmath,amsfonts}
\usepackage{algorithmic}
\usepackage{algorithm}
\usepackage{array}
\usepackage[caption=false,font=normalsize,labelfont=sf,textfont=sf]{subfig}
\usepackage{textcomp}
\usepackage{stfloats}
\usepackage{url}
\usepackage{verbatim}
\usepackage{graphicx}
\usepackage{cite}
\usepackage{multirow}
\usepackage{booktabs}
\begin{document}

\title{EsaNet: Environment Semantics Enabled Physical Layer Authentication}

\author{Ning Gao,~\IEEEmembership{Member,~IEEE,}~Qiying Huang,~Cen Li,~Shi Jin,~\IEEEmembership{Senior Member,~IEEE,}\\and~Michail Matthaiou,~\IEEEmembership{Fellow,~IEEE}


\thanks{The work of M. Matthaiou was supported in part by the European Research Council (ERC) under the European Union's Horizon 2020 research and innovation programme (grant agreement No. 101001331).} 
\thanks{N. Gao, Q. Huang and Cen Li are with the School of Cyber Science
and Engineering, Southeast University, Nanjing 210096, China (e-mail:
ninggao@seu.edu.cn; huangqiying@seu.edu.cn).}
\thanks{S. Jin is with the National Mobile Communications
Research Laboratory, Southeast University, Nanjing 210096, China (e-mail: jinshi@seu.edu.cn).}
\thanks{
M. Matthaiou is with the Centre for Wireless Innovation (CWI), Queen’s University Belfast, Belfast BT3 9DT, U.K. (e-mail: m.matthaiou@qub.ac.uk).}
}

%
%
\maketitle

\begin{abstract}
Wireless networks are vulnerable to physical layer spoofing attacks due to the wireless broadcast nature, thus, integrating communications and security (ICAS) is urgently needed for 6G endogenous security. In this letter, we propose an environment semantics enabled physical layer authentication network based on deep learning, namely EsaNet, to authenticate the spoofing from the underlying wireless protocol. Specifically, the frequency independent wireless channel fingerprint (FiFP) is extracted from the channel state information (CSI) of a massive multi-input multi-output (MIMO) system based on environment semantics knowledge. Then, we transform the received signal into a two-dimensional red green blue (RGB) image and apply the you only look once (YOLO), a single-stage object detection network, to quickly capture the FiFP. Next, a lightweight classification network is designed to distinguish the legitimate from the illegitimate users. Finally, the experimental results show that the proposed EsaNet can effectively detect physical layer spoofing attacks and is robust in time-varying wireless environments.
\end{abstract}


\section{Introduction}
\IEEEPARstart{W}{ireless} networks are highly vulnerable to the threats of physical layer spoofing attacks due to the broadcast nature of wireless communications. For example, in IEEE 802.11n Wi-Fi network, a spoofer can forge the media access control (MAC) address of a legitimate user through the $\mathrm{ifconfig}$ command to access the Wi-Fi network, and can further launch a series of attacks, such as session hijacking, etc., \cite{7109166}. Unfortunately, the high-level authentication mechanisms have difficulties in detecting such physical layer attacks. Physical layer authentication is based on the communication subsidiary products to distinguish the user's identity, which is wireless endogenous and unique to any particular user \cite{7539590}. The commonly used endogenous products are channel fingerprints, such as the channel impulse response (CIR), channel frequency response (CFR), received signal strength (RSS) or radio frequency (RF) fingerprints \cite{9450821}, such as carrier frequency offset (CFO) and input/output (I/Q) imbalance, etc.

Generally, most of the existing physical layer authentication methods include two categories according to the usage of the threshold, which include fixed threshold, dynamic threshold and threshold-free methods \cite{9279294}. The fixed threshold methods usually have to calculate the corresponding target detection probability, which overly requires strong assumptions and prior knowledge of the wireless environments \cite{7839273,4533330}. However, with the development of B5G/6G, wireless electromagnetic environments are becoming increasingly complex and cannot be modeled by classical channel models, especially for higher frequencies, such as mmWave and THz \cite{Papasotiriou}. Thus, the accurate model parameters are difficult to obtain and the statistical properties of the channel do no longer follow a known distribution, so calculating an optimal detection threshold is impossible. Thanks to the strong nonlinear learning ability of artificial intelligence (AI), the AI enabled dynamic threshold/ threshold-free methods, such as Bayesian classifiers, extreme learning machines, deep learning and reinforcement learning \cite{8961122,GAO,7891506}, are gradually attracting increasing research attention. These methods can achieve a satisfactory performance in unknown wireless environments. In addition, some works have transformed the channel fingerprint or RF fingerprint matrix into an image to reduce the data processing overhead and accelerate the network training \cite{8322184,9790811,8979381,9517121}. The authors of \cite{8979381} proposed an adaptive neural network to track the time-varying channel state information (CSI) image and realize intelligent authentication. The work of \cite{9517121} proposed an end-to-end
deep learning scheme for extracting the RF fingerprint to achieve authentication of the different transmitters.

However, due to upcoming versions of massive multiple-input multiple-output (MIMO) architectures, the complexity of CSI acquisition will increase exponentially \cite{9931713}. One significant issue is that directly using the CSI to detect the spoofing attack leads to a high dimensional data processing overhead, followed by a long authentication latency. This is disastrous for delay-sensitive wireless networks. Moreover, directly inputting CSI into neural networks compromises data interpretability, whilst the inefficient network training results in unstable authentication performance in time-varying wireless environments. To the best of the authors' knowledge, it is the first time that the environment semantics knowledge of the wireless channel is proposed for physical layer authentication. In this letter, by analyzing the double time-scale features of the massive MIMO channel, we filter out the fast time-varying features and select the frequency independent features as the wireless channel fingerprint (FiFP) for spoofing detection. Henceforth, we propose a robust environment semantics enabled physical layer authentication network, namely EsaNet. Our contributions are summarized as follows:

\begin{itemize}
\item{Different from existing works, in which the neural network is regarded as a black-box and the CSI is fed straightly into the neural network, for the first time, based on the environment semantics knowledge, we extract the angle-delay features as FiFP, which has a higher interpretability and is efficient for neural network training. With the trained EsaNet, the physical layer spoofing attack can be effectively detected in time-varying wireless environments.}
\item{For the proposed EsaNet, we use the you only look once (YOLO), an advanced single-stage object detection network, to quickly capture the angle-delay features from the received signal, which can significantly reduce the data processing overhead and the authentication latency. Next, a lightweight neural network is developed to conduct the classification without a detection threshold, which is suitable for unknown wireless environments.}
\item{The simulations show that the proposed EsaNet can effectively authenticate users and is robust in time-varying wireless environments. In addition, it is compatible with the existing communication protocols and shares components with the channel estimation process. This paradigm of simultaneous transmission and identity authentication has great potential for integrating communications and security (ICAS) at 6G endogenous security \cite{ARXIV22GAO}.}
\end{itemize}

\section{System Model}
\begin{figure}[!t]
\centering
\includegraphics[width=2.5in]{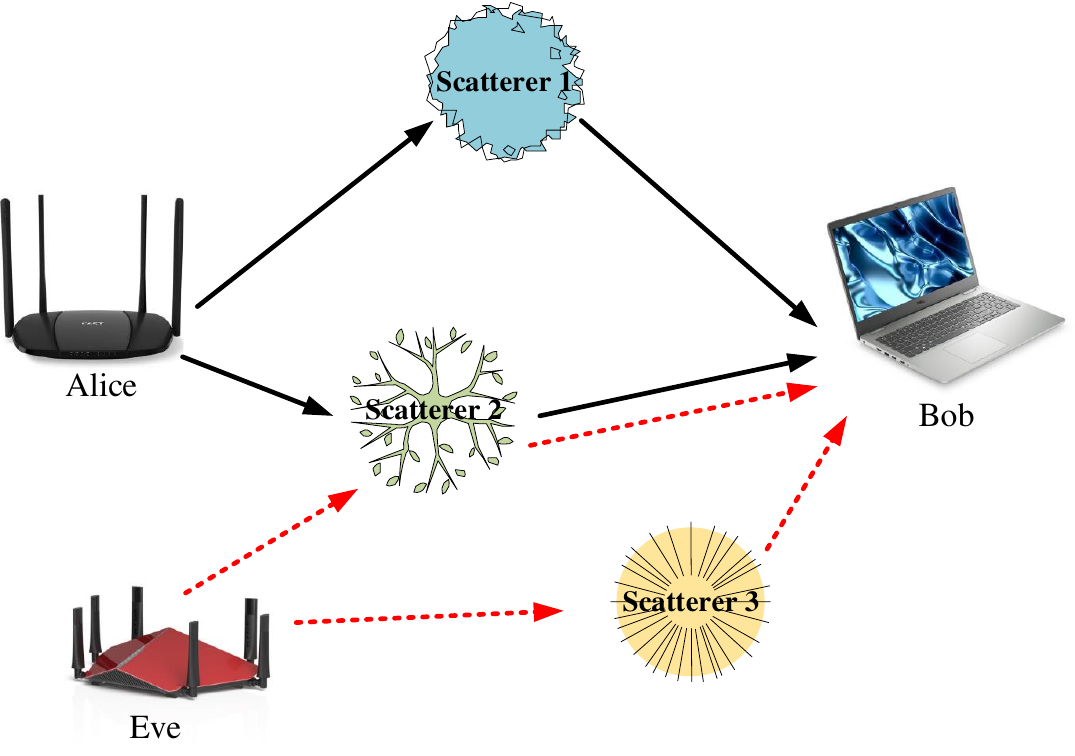}
\caption{Schematic diagram of the system model.}
\label{fig1}
\end{figure}

We consider a typical three users indoor scenario in Fig. \ref{fig1}, where the legitimate transmitter Alice communicates with the legitimate receiver Bob in a multi-scatter environment, while the malicious user Eve tries to access Bob by imitating Alice's identity. We assume that Alice, Bob and Eve are geographically located at different positions and each location is fixed. We adopt a MIMO orthogonal frequency division multiplexing (MIMO-OFDM) downlink communication system, where both Alice and Eve are equipped with a uniform linear array (ULA) antenna array with $M$  antennas separated by interval $d$ while Bob is equipped with a single antenna. Each frequency band includes $N$ subcarriers with spacing $\Delta{f}$ and the center frequency is $f_0$ with bandwidth $W$.

The geometric channel across all the antennas and subcarriers between Alice and Bob can be modeled as
\begin{equation}
\label{deqn_ex1a}
{\mathbf H}_u = \sum_{l=1}^{L} g_{l} {\mathbf a}_u(\Theta_{l}) {\mathbf p}_u^T(\Gamma_{l}),
\end{equation}

\noindent where ${\mathbf H}_u\in \mathbb{C} ^{M\times N}$ with $u\in \left \{ \rm Alice, Bob \right \}$ or $\left \{ \rm Eve, Bob \right \}$, $L$ is the number of propagation paths and $g_l$ denotes the complex gain of the $l$th path. The steering vector of the ULA and the delay-related phase vector of subcarriers are expressed as
\begin{equation}
\label{dx2}
{\mathbf a}_u(\Theta_{l}) = \left[1, e^{j 2 \pi \Theta_{l}}, \ldots, e^{j 2 \pi(M-1) \Theta_{l}}\right]^T,
\end{equation}
and
\begin{equation}
\label{deqn_ex1a}
{\mathbf p}_u(\Gamma_{l}) = \left[1, e^{j 2 \pi \Gamma_{l}}, \ldots, e^{j 2 \pi(N-1) \Gamma_{l}}\right]^T,
\end{equation}

\noindent respectively, where $\Theta_{l}= \frac{d}{\lambda }\sin \theta_{l} \in \left [ 0,1 \right ] $, $d = \frac{1}{2}\lambda $, $\theta_{l}$ denotes the angle of the $l$th path; $\Gamma_{l}=\Delta{f}\tau_{l} \in \left [ 0,1 \right )$, where $\tau_{l}$ denotes the delay of the $l$th path.

By assuming an all-ones transmitted pilot across all antennas and subcarriers, the signal at the receiver is denoted as
\begin{equation}
    \begin{aligned}
\label{deqn_ex1a}
{\mathbf Y}_u = \sqrt{P}{\mathbf H}_u+ {\mathbf Z},
    \end{aligned}
\end{equation}

\noindent where $\mathbf{Y}_u\in \mathbb{C} ^{M\times N}$, $P$ is the transmitting power and ${\mathbf Z}\in \mathbb{C} ^{M\times N}$ denotes the additive Gaussian white noise. Without loss of generality, in the following analysis, we use the symbol $\mathbf{Y}$ to represent the received signal from both Alice and Eve.

\textit{Remark}: It indicates that CSI is unique for any transceiver pair in different locations. Compared to the fast time-varying gain and phase, the scatterers corresponding to angle-delay characteristics are frequency independent, sparse and slow-varying (FiFP), and, for these reasons, are regarded as the robust environment semantics feature.

\section{Environment Semantics Aided Physical Layer Authentication Scheme}
\begin{figure}[!t]
\centering
\includegraphics[width=3.4in]{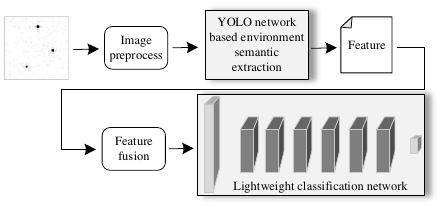}
\caption{The framework of the proposed EsaNet.}
\label{fig2}
\end{figure}

In this section, we first derive the angle-delay image, and then design the framework of the proposed EsaNet, which is shown in Fig. \ref{fig2}. After that, we present the complete algorithm.
\subsection{Angle-Delay Image Generation}
In a massive MIMO-OFDM communication system, the number of paths is typically much smaller than the number of transmitter antennas and subcarriers, i.e., $L \ll  M, N $. Therefore, the MIMO channel is sparse when it is transformed from the antenna-subcarrier domain to the angle-delay domain. Therefore, the received ${\mathbf Y}$ can be mapped to a sparse $\overline{{\mathbf Y}}$ through an angle base ${\mathbf U}_\Theta$ and a delay base ${\mathbf U}_\Gamma$, which is given by
\begin{equation}
\label{deqn_ex1a}
\overline{{\mathbf Y}}= {\mathbf U}_\Theta^T {\mathbf Y} {\mathbf U}_\Gamma ,
\end{equation}
and
\begin{equation}
\label{deqn_ex1a}
{\mathbf U}_\Theta = \left[{\mathbf a}(0),{\mathbf a}\bigg(-\frac{1}{\alpha M }\bigg), \ldots, {\mathbf a}\bigg(-\frac{\alpha M - 1}{\alpha M }\bigg)\right] ,
\end{equation}
\begin{equation}
\label{deqn_ex1a}
{\mathbf U}_\Gamma = \left[{\mathbf p}(0),{\mathbf p}\bigg(-\frac{1}{\beta N }\bigg), \ldots, {\mathbf p}\bigg(-\frac{\beta N - 1}{\beta N }\bigg)\right] ,
\end{equation}

\noindent where $\overline{{\mathbf Y}} \in \mathbb{C}^{\alpha M \times \beta N }$, $\alpha$ and $\beta$ are oversampling rates respectively. The $\left ( m,n \right )$th entry of $\overline{{\mathbf Y}}$ is expressed as
\begin{equation}
\label{deqn_ex1a}
\left [ \overline{{\mathbf Y}} \right ]_{m,n} = \sum_{l=1}^{L} \epsilon_{l} k_{\Theta,l} k_{\Gamma,l},
\end{equation}

\noindent where
\begin{equation}
\label{dx9}
k_{\Theta,l} = {\mathbf a}^H\bigg(-\frac{m}{\alpha M }\bigg) {\mathbf a}(\Theta_l),
\end{equation}

\noindent and
\begin{equation}
\label{dx10}
k_{\Gamma,l} = {\mathbf p}^T(\Gamma_{l}) {\mathbf p}\bigg(-\frac{n}{\beta N}\bigg).
\end{equation}
By invoking Eq. \eqref{dx2} and utilizing the property of geometric progression, \eqref{dx9} can be rewritten as
\begin{equation}
\label{dx11}
        k_{\Theta,l} = \frac{1-e^{j2\pi M(\Theta_l+\frac{m}{\alpha M })}} {1-e^{j2\pi (\Theta_l+\frac{m}{\alpha M })}}.
\end{equation}
According to the mathematical relationship
\begin{equation}
\label{deqn_ex1a}
1-e^{j2\pi(\Theta_l+\frac{m}{\alpha M })} = e^{j\pi M(\Theta_l+\frac{m}{\alpha M })}\sin\big(\pi (\Theta_l+\frac{m}{\alpha M })\big),
\end{equation}
the module of Eq. \eqref{dx11} can be given by
\begin{equation}
\label{deqn_ex1a}
\left | k_{\Theta,l} \right | = \frac{\sin\big(\pi M(\Theta_l+\frac{m}{\alpha M })\big)}{\sin\big(\pi(\Theta_l+\frac{m}{\alpha M })\big)},
\end{equation}
which attains its maximum value when
\begin{equation}
\label{deqn_ex1a}
\Theta_l + \frac{m}{\alpha M } = 0.
\end{equation}
Similarly, the module of \eqref{dx10} can be represented as
\begin{equation}
\label{deqn_ex1a}
\left | k_{\Gamma,l} \right | = \frac{\sin\big(\pi N(\Gamma_l - \frac{n}{\beta N })\big)}{\sin\big(\pi(\Gamma_l -\frac{n}{\beta N })\big)},
\end{equation}

\noindent which is maximum when the following condition is satisfied
\begin{equation}
\label{deqn_ex1a}
\Gamma_l - \frac{n}{\beta N } = 0.
\end{equation}
Afterwards, a new matrix $\widetilde{{\mathbf Y}}$ is obtained through normalizing $\overline{{\mathbf Y}}$ to the maximum value $\delta$, such that
\begin{equation}
\label{deqn_ex1a}
\left [ \widetilde{{\mathbf Y}} \right ]_{m,n} = \frac{\delta}{\max\limits_{i=1, \ldots, \alpha M \atop j=1, \ldots, \beta N}{\left |\left [ \overline{{\mathbf Y}} \right ]_{i,j}\right | }} \left |\left [ \overline{{\mathbf Y}}\right ]_{m,n} \right |,
\end{equation}

\noindent and the RGB image of $\widetilde{{\mathbf Y}}$ can be drawn by Pcolor function.

\textit{Remark}: We Assume that the image has $L$ heat spots corresponding to the environment semantics. The vertical coordinate of the $l$th heat spot center is the angle of the $l$th path and the horizontal coordinate of the $l$th heat spot center is the delay of the $l$th path, which are $\left | \Theta_l \right | = m / (\alpha M)$  and $\left | \Gamma_l \right | = n / (\beta N)$, respectively. Furthermore, we can calculate the center of the $l$th heat spot to estimate $\Theta_l$ and $\Gamma_l$, which ranges from 0 to 1. To accurately determine the location of the bounding box, the value of matrix $\widetilde{{\mathbf Y}}$, within the range of the bounding box, is not below 1/20 of its peak value.

\subsection{YOLO Based Feature Capture}
To quickly and accurately capture the environment semantics in each image, a YOLOv5 network is applied. Specifically, the backbone network adopted by YOLOv5 is CSPDarknet53, which adds a cross stage partial network to each large residual block in Darknet53. Furthermore, YOLOv5 utilizes a feature pyramid network and a path aggregation network to achieve target feature extraction of different sizes.

To train the network, we set the label of the $l$th heat spot in each image as $\left({x_{min}, y_{min}, x_{max}, y_{max}, {\rm class}} \right)$, where class is set to be 0. A YOLOv5 network can capture the bounding box from the upper left vertex to the lower right vertex of the heat spot, i.e., $\left ( {x_{min}, y_{min}} \right )$ and $\left ( {x_{max}, y_{max}} \right )$. In this case, the angle and the delay of the $l$th path can be calculated by
\begin{equation}
\label{deqn_ex1a}
\hat{\Theta}_l = 1 - \frac{y_{lmin}+y_{lmax}}{2\times 938},
\end{equation}
and
\begin{equation}
\label{deqn_ex1a}
\hat{\Gamma}_l = \frac{x_{lmin}+x_{lmax}}{2\times 938}.
\end{equation}

\begin{algorithm}
\caption{~~~The application pipeline of EsaNet}\label{alg:1}
\begin{algorithmic}[1]
\STATE \textbf{Initialization:} Parameters in YOLOv5 and classification network.
\STATE \textbf{Offline training:}
\FOR {each received signal $\mathbf Y$}
\STATE Transform $\mathbf Y$ into an image and build the image set;
\ENDFOR
\STATE Divide the image set into the training and validation sets;
\STATE Input the training set into YOLOv5 network for training;
\FOR {Each classification image}
\STATE Capture the angle and the delay of each path $\Theta_l$ and $\Gamma_l$ with the trained YOLOv5 network;
\STATE Fuse feature and input it into the lightweight classification network for training;
\ENDFOR
\STATE \textbf{Online authentication:}
\FOR {Each validation image}
\STATE Classify the transmitter with the trained EsaNet.
\ENDFOR
\end{algorithmic}
\end{algorithm}

\textit{Remark}: The EsaNet learns the identity classification based on statistical property rather than specific values, thus the trained EsaNet can also be applied in multi-user authentication.

\subsection{Lightweight Classification Network}
As the cascade structure of the EsaNet, the developed classification network adopts four Dense layers to extract features and add a Dropout layer after every two Dense layers to prevent over-fitting. The $\mathrm{RELU}$ function is used in the third layer and $\mathrm{sigmod}$ function is applied in the other layers.

The loss function for model training is the cross-entropy loss function, which is expressed by
\begin{equation}
\label{deqn_ex1a}
L=-[c \log \hat{c}+(1-c) \log (1-\hat{c})],
\end{equation}

\noindent where the classification label of the legitimate user and the illegitimate user are set to 1 and 0, respectively. The symbol $c$ is the label of the data and $\hat {c}$ is the probability that the model predicts the data coming from the legitimate user.

\subsection{Complete Algorithm of EsaNet}
The proposed algorithm can be divided into two stages, which includes offline training and online authentication. In offline training, the dataset is transformed into a RGB image, and then the generated images are divided into two categories, i.e., the training set and the validation set, which are used for training and validating the YOLOv5 network, respectively. Afterwards, the angle-delay features are captured using the trained YOLOv5 network. The output features of YOLOv5 network are fused and fed into the lightweight classification network for training. In online authentication, the receiver authenticates the transmitter based on the trained EsaNet.

\section{Simulations}
\subsection{Dataset Generation and Network Training}
The experimental configuration is a Pytorch framework for NVIDIA GeForce GTX 2080Ti. The YOLOv5 network parameter sets are trained by Adam optimizer to fine-tune the network. Considering that the pixel values of RGB images range from 0 to 255, the maximum value $\delta$ is set to be 255. To train the YOLOv5 network, by setting different signal-to-noise ratio (SNR), we generate 1800 channel images and corresponding labels with the angle-delay following a $\mathcal{N}\sim(0,1)$. To train and verify the classification network, we generate dataset A and dataset B based on our scenario. For dataset A, the angle-delay follow normal distribution with a fixed mean value and variance, where Alice follows $\mathcal{N}\sim(0.3,0.1)$ and Eve follows $\mathcal{N}\sim(0.6,0.1)$, respectively. For dataset B, the angle-delay follow normal distribution but with different mean values, where the 50\% data of Alice follows $\mathcal{N}\sim(0.3,0.1)$ and 50\% data of Alice follows $\mathcal{N}\sim(0.4,0.1)$, meanwhile, 50\% data of Eve follows $\mathcal{N}\sim(0.6,0.1)$ and 50\% data of Eve follows $\mathcal{N}\sim(0.7,0.1)$, respectively. We use the dataset A to train and validate the performance of the proposed EsaNet, and utilize the dataset B to analyze the robustness of the proposed EsaNet in time-varying wireless environments.

\begin{table}[!h]
\caption{Parameter values.}
\label{Tab:table1}
\centering
\renewcommand\arraystretch{1}
\setlength{\tabcolsep}{4mm}{
\begin{tabular}{ccc}
\toprule
Setting & Parameter & Value\\
\midrule
\multirow{8}{*}{Channel environment} & $M$ & 32\\
& $N$ & 32\\
& $f_0$ & 3.5GHz\\
& $\Delta f$ & 60kHz\\
& $L$ & 5\\
& $\alpha,\beta$ & 16\\
& SNR & $[0,25]$dB\\
& $\delta$ & 255\\
\midrule
\multirow{6}{*}{YOLOv5 network} & Learning rate & $1 \times 10^{-3}$\\
& Weight decay coefficient & $5 \times 10^{-4}$\\
& Number of iterations & 300\\
& Batch size & 16\\
& Confidence threshold & 0.25\\
& IOU threshold & 0.45\\
\midrule
\multirow{3}{*}{Classification network} & Learning rate & $1 \times 10^{-3}$\\
& Number of iterations & 1200\\
& Batch size & 64\\
\bottomrule
\end{tabular}}
\end{table}

The YOLOv5 network is trained to capture the angle-delay features, and then the lightweight classification network is trained. To train this network, we get a set with 300 angle and delay values with precision of four decimal places detected by the trained YOLOv5 network. Regarding the detection error, we adopt the first three propagation paths, the length and width information of the bounding box as the fusion feature, which is the input of the classification network. The classification network is trained by the root mean square prop optimizer, and the classification accuracy is adopted to evaluate the network. The main parameter settings are shown in Table I.

\subsection{Performance Analysis}
\begin{figure}[!t]
\centering
\includegraphics[width=3.25in]{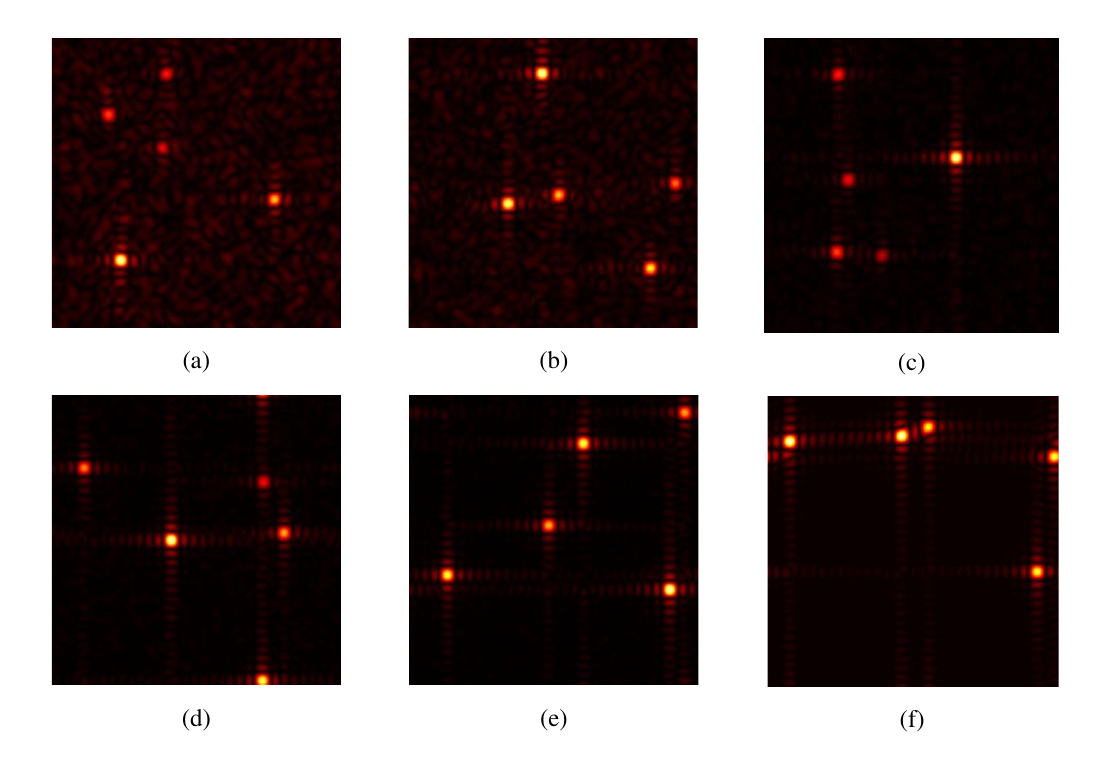}
\caption{The angle-delay images in different SNR conditions $ \rm (a)~SNR = 0~dB\ (b)~SNR = 5~dB\ (c)~SNR = 10~dB\ (d)~SNR = 15~dB\ (e)~SNR = 20~dB\ (f)~SNR = 25~dB$.}
\label{fig_3}
\end{figure}

\begin{figure}[!t]
\centering
\includegraphics[width=3.1in]{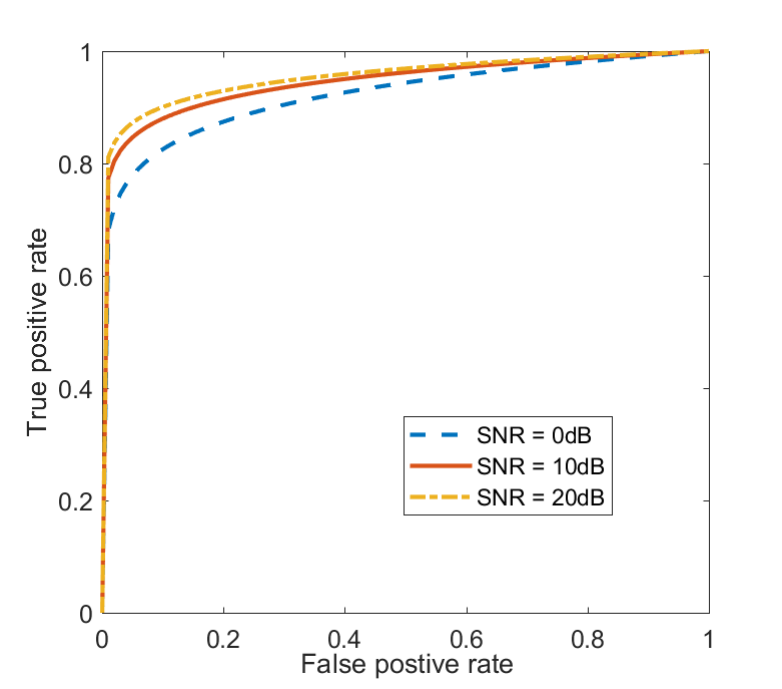}
\caption{The ROC curve in different SNR conditions.}
\label{fig_4}
\end{figure}

Figure 3 shows the feature capture performance of the trained YOLOv5 network, illustrated through the angle-delay images in different SNRs. It can be observed that the heat spots are relatively large and show regular squares, while the noise spots show no regularity. With the increase of the SNR, the noise spots decrease, and the outline of the angle-delay spots become brighter. Particularly, when the value of SNR is 10 dB and above, the noise levels have limited influence on the heat spots' detection.

\begin{table}[!h]
\caption{Detection accuracy of different methods for dataset A}
\label{Tab:table2}
\centering
\renewcommand\arraystretch{1.1}
\begin{tabular}{c|c|c|c|c|c}
\hline
SNR  & 0dB & 5dB & 10dB & 15dB & 20dB\\
\hline
\textbf{EsaNet} & \textbf{99.25} & \textbf{99.5} & \textbf{99.75} & \textbf{99.75} & \textbf{99.75}\\
\hline
VGG-net & 88.75 & 92 & 91.5 & 90.75 & 89.75\\
\hline
Fixed threshold & 97.5 & 96.8 & 97.7 & 97.3 & 97.5\\
\hline
\end{tabular}
\end{table}

\begin{table}[!h]
\caption{Detection accuracy of different methods for dataset B}
\label{Tab:table3}
\centering
\renewcommand\arraystretch{1.1}
\begin{tabular}{c|c|c|c|c|c}
\hline
SNR  & 0dB & 5dB & 10dB & 15dB & 20dB\\
\hline
\textbf{EsaNet} & \textbf{92.5} & \textbf{92.7} & \textbf{92.6} & \textbf{93.5} & \textbf{92.5}\\
\hline
VGG-net & 84 & 81 & 80 & 81 & 80\\
\hline
Fixed threshold & 90.3 & 91.5 & 90 & 90.4 & 90.8\\
\hline
\end{tabular}
\end{table}

In Fig. 4, the receiver operating characteristic (ROC) curves under different SNRs are used to evaluate the detection performance. A larger area under the ROC curve means a better performance of the classification. Thus, from the figure, we can find that the ROC curves are very close to each other, which shows that the proposed network has robustness at different noise levels. In addition, we can observe that the areas under the ROC curves are 0.9925, 0.9975, 0.9985 for 0 dB,10 dB and 20 dB, respectively, which shows that the proposed EsaNet performs well in identity classification between the legitimate user and the illegitimate user.

We now evaluate the detection accuracy of different methods under various SNRs in Table II, where the fixed threshold detection and VGG-net based black-box detection are considered as benchmarks. The table shows that the proposed method can achieve a consistently robust performance with the increase of SNR, which is above 99\%, and is robust to different noise environments. For the benchmarks, the VGG-net is significantly affected by the noise level. For example, we can observe that with different SNRs, the VGG-net has nearly 4\% performance fluctuation while our proposed EsaNet has only 0.5\% performance fluctuation. In addition, the proposed EsaNet directly analyzes the received signal and eliminates the detection errors via feature fusion, which has low overhead and is not affected by estimation errors. However, the VGG-net based black-box detection uses the CSI as the input; consequently, its detection accuracy is influenced by the channel estimation error. The larger the channel estimation error, the lower the detection accuracy. The detection accuracy of the fixed threshold is relatively stable (above 96.8\%). Notably, although the fixed threshold detection can achieve a desirable detection accuracy in different SNRs, the optimization detection threshold is hard to obtain in dynamic unknown wireless environments. In this case, we can also conjecture that the detection accuracy of the fixed threshold detection is lower than that of the proposed EsaNet. Next, based on the trained EsaNet, the VGG-net and the optimization fixed threshold, we utilize dataset B to analyze the detect robustness of the methods in different scattering environments. Table III shows that with the mean value change of the angle-delay distribution, the detection accuracy decreases for all three methods. The detection accuracy reduction of the proposed EsaNet, the VGG-net and the fixed threshold is 6.85\%, 9.35\% and 6.76\%, respectively, which suggests that our proposed EsaNet is robust. The reason is that our proposed approach removes the nonlinear effect of the fast-varying channel gain and semanticizes the change of the wireless environments as the linear translation of the heat spots in an image.

\section{Conclusion}
This letter has proposed a rubust environment semantics enabled physical layer authentication network, namely EsaNet. Specifically, the environment semantics knowledge has been used to extract the angle-delay features as FiFP, which enhances the interpretability and environment robustness. Then, we have utilized the YOLOv5 network to quickly capture the features
and designed the lightweight classification network to authenticate the users without a detection threshold. Simulations show that the proposed EsaNet has a good detection accuracy and is robust in time-varying wireless environments.

\ifCLASSOPTIONcaptionsoff
  \newpage
\fi
\bibliographystyle{IEEEtran}
\bibliography{IEEEabrv,sigproc} 

\end{document}